\documentclass[aps,pre,twocolumn,showpacs,superscriptaddress,floatfix]{revtex4-1} 
\usepackage{amsmath}
\usepackage{amsfonts}
\usepackage{amssymb}
\usepackage{graphicx}

\graphicspath{{./}{figures/}}

\begin{document}

\title{Collective Motion and Phase Transitions of Symmetric Camphor Boats}

\author{Eric Heisler}
\affiliation{Department of Mathematical and Life Sciences, Hiroshima University, 1-3-1 Kagamiyama, Higashi-Hiroshima 739-8526, Japan}
\author{Nobuhiko J. Suematsu}
\affiliation{Graduate School of Advanced Mathematical Sciences, Meiji University, 1-1-1 Higashimita, Tamaku, Kawasaki 214-8571, Japan}
\affiliation{Meiji Institute for Advanced Study of Mathematical Sciences (MIMS), 1-1-1 Higashimita, Tamaku, Kawasaki 214-8571, Japan}
\author{Akinori Awazu}
\affiliation{Department of Mathematical and Life Sciences, Hiroshima University, 1-3-1 Kagamiyama, Higashi-Hiroshima 739-8526, Japan}
\author{ Hiraku Nishimori}
\affiliation{Department of Mathematical and Life Sciences, Hiroshima University, 1-3-1 Kagamiyama, Higashi-Hiroshima 739-8526, Japan}

\date{\today}

\begin{abstract}
The motion of several self-propelled boats in a narrow channel displays spontaneous pattern formation and kinetic phase transitions. In contrast with previous studies on self-propelled particles, this model does not require stochastic fluctuations and it is experimentally accessible. By varying the viscosity in the system, it is possible to form either a stationary state, correlated or uncorrelated oscillations, or unidirectional flow. Here, we describe and analyze these self organized patterns and their transitions.
\end{abstract}

\maketitle

\section{Introduction}
Groups of motile, interacting bodies are observed everywhere in nature, from groups of animals\cite{Couzin2003, Buhl2006, Vicsek2010} to mechanical systems~\cite{Blair2003, Deseigne2010}. To better understand and unify this broad category of collective motion, it is beneficial to construct and analyze simpler mathematical models. One particularly useful, yet simple model is that of Self-Propelled Particles(SPPs) as first described by Vicsek and Zafiris~\cite{Vicsek1995}. Although the original SPP models are very simple, they have displayed a variety of complex behaviors including kinetic phase transitions~\cite{Vicsek1995, Czirok1999, Levine2000, Gregoire2004} and large scale pattern formation~\cite{Czirok1997}. More specific versions have also been made, which exhibit similar behaviors~\cite{Bertin2006, Dorsogna2006, Baskaran2008, Mishra2010}. The benefit of a more specific model is greater applicability to physical systems. Unfortunately, it is often difficult to test the results of SPP models because the relevant parameters cannot be controlled experimentally. The model presented here shares many similarities with SPPs, but has the great advantage of being experimentally accessible. It displays many of the interesting behaviors seen in simpler models as well as new kinds of self organization. There is no need to introduce artificial fluctuations, and the important parameters are relatively easy to control in experiment.

This model describes the motion of an ensemble of symmetric camphor boats(CBs) floating on water. CBs are propelled by surface tension gradients generated by attached pieces of camphor. They interact, aside from collisions, only through their influence on the background camphor field. Two-dimensional experiments with symmetric camphor particles~\cite{Soh2008} have displayed spontaneous pattern formation which appears very similar to results of 2-D simulations using this model. However, their analysis was based on hydrodynamic effects rather than surface tension. Ensembles of asymmetric CBs have been studied experimentally~\cite{Nakata2005, Suematsu2010}, but are different in that they have a fixed orientation and are driven in a specified direction. This paper focuses on the results of 1-D simulations and analysis, and some 2-D results are included. In the future we hope to test this model through experiment in both one and two dimensions.

There are three major differences between CBs and basic SPPs. First, SPPs always move at a constant speed. In this model, and some others~\cite{Bertin2006, Dorsogna2006, Baskaran2008, Mishra2010}, boats are subject to various forces. However, there is a characteristic speed corresponding to a free flowing boat. Second, The direction of travel of SPPs is instantaneously set depending on neighboring particles. CBs are influenced by collisions with other boats and by a surface tension gradient. The latter is determined by nearby boats and the history of that region. Third, there are no random fluctuations intentionally added to this system. Most other SPP models depend on added noise, and some use it as a key parameter.

In numerical simulations of the CB model, we observed two distinct kinetic phase transitions by varying the viscosity of the water. One appears as the abrupt formation of a stationary, ordered pattern, and the other as a discontinuity in the mean velocity, $\lvert \left<v\right>\rvert$, of the ensemble. Aside from these transitions, a variety of collective behaviors were observed such as synchronized formations and erratic oscillations. Here we will describe some of these behaviors and present quantitative evidence for the phase transitions and pattern formation.

\begin{figure}[hbt]
	\includegraphics[width=8cm]{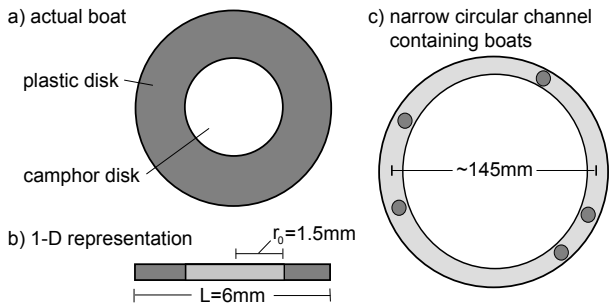}
	\caption{A schematic showing $a)$an actual boat in 2-D and $b)$the 1-D representation used here. $c)$ shows the periodic route in 2-D.}
	\label{fig:aboat}
\end{figure}
\section{Model}
The model being considered describes the motion of thin plastic disks which float on the surface of water. Smaller camphor pellets are attached to the center of the underside of the disks. A schematic illustration is shown in Fig.(\ref{fig:aboat}). As the camphor dissolves and diffuses in the water, it changes the surface tension. The resulting tension gradients propel the boats. Such systems have often been modeled by a simple set of equations based on surface tension and viscosity~\cite{Nagayama2004, Suematsu2010, Kohira2001, Hayashima2001}. This version was also used by us in ~\cite{Heisler2011} and these CBs are similar to those described in~\cite{Suematsu20102}, but are constructed symmetrically, so there is no preferential direction of travel. They move in a circular channel with circumference $R$, which is narrow enough to restrict the motion to one dimension. The 1-D equations of motion are given by (\ref{eq:model_v1}). These equations can be readily extended to 2-D by vectorizing them and integrating the surface tension around the edge of the boat.
\begin{equation}
\frac{\partial^{2}x}{\partial t^{2}} = -\frac{\mu}{m} \frac{\partial x}{\partial t} + \frac{L}{m}\left[ \gamma (c(x+L/2)) - \gamma (c(x-L/2))\right]
\label{eq:model_v1}
\end{equation}
where $m$ is the mass of the boat, $\mu$ is the viscosity constant of the water and $L$ is the length of the boat. The position, $x$, representing the center of the boat, is defined on a periodic domain with period $R$. The second term on the right represents the difference in surface tension between the front and back of the boat as a function of the camphor concentration given by $c(x+L/2)$ and $c(x-L/2)$ respectively. $\gamma(c)$ is approximated by the sigmoidal function in eq.\ref{eq:model_gamma}.
\begin{equation}
\gamma(c) =\frac{\gamma_{water} - \gamma_{camphor}}{\left(\beta c\right)^{2} +1} + \gamma_{camphor}
\label{eq:model_gamma}
\end{equation}
$\gamma_{water}$ and $\gamma_{camphor}$ are the surface tension of pure water and camphor saturated solution respectively.

The concentration of camphor molecules on the surface of the water is constantly changing due to several processes. For a system with $N$ boats, it can be approximated by the following reaction-diffusion equation (\ref{eq:model_c}).
\begin{eqnarray}
\frac{\partial c}{\partial t} = D\frac{\partial^{2}c}{\partial x^{2}} - kc + \alpha \sum_{i=1}^{N} F(x-x_{i}) \label{eq:model_c} \\
F(x) = 1 \ \rm{: for} \ \left| x \right| \leq r_{0}, \ 0 \ \rm{: otherwise} \nonumber
\end{eqnarray}
Here, $D$ is the diffusion constant, $k$ is a constant combining the effects of evaporation and dissolution, and $\alpha F(x-x_{i})$ represents the addition of camphor by each boat's pellet, which is centered at the point $x_{i}$ and has half-length $r_{0}$. To non-dimensionalize the problem, we define the following dimensionless quantities.
\begin{equation}
t'= t \frac{D}{L^{2}} \ , \ x'=\frac{x}{L} \ , \ c'= c \beta
\label{eq:nondimquant}
\end{equation}
Then the dimensionless parameters of the system become
\begin{eqnarray}
\mu'=\frac{\mu L^{2}}{m D} \ , \ k' = \frac{k L^{2}}{D} & , \ & \Gamma=\frac{L^{3}(\gamma_{w} - \gamma_{c})}{m D^{2}} \\ \nonumber
r_{0}' = \frac{r_{0}}{L} \ , \ R' = \frac{R}{L} & , \ & \alpha'=\frac{\alpha \beta L^{2}}{D}
\label{eq:nondimparam}
\end{eqnarray}
Dropping the $'$ marks, the non-dimensional equations are
\begin{eqnarray}
\frac{\partial^{2}x_{i}}{\partial t^{2}} & = & - \mu \frac{\partial x_{i}}{\partial t}  + \Gamma \left[ \frac{1}{c(x_{i}+\frac{1}{2})^{2} + 1} - \frac{1}{c(x_{i}-\frac{1}{2})^{2} + 1} \right]  \nonumber \\
\frac{\partial c}{\partial t} & = & \frac{\partial^{2}c}{\partial x^{2}} - kc + \alpha \sum_{i=1}^{N} F(x-x_{i}) \label{eq:NDcamphor} \\
\ & \ & F(x) = 1 \ \rm{: for} \ \left| x \right| < r_{0}, \ 0 \ \rm{: otherwise} \nonumber 
\end{eqnarray}

To approximate a realistic system, we used parameters corresponding to the dimensional values: $R=45.5cm$, $L=0.6cm$, $r_{0}=0.15cm$, $m=0.009g$, $\gamma_{w}=72g/s^{2}$, $\gamma_{c}=50g/s^{2}$, $D=1cm^{2}/s$ (other parameters being varied). Also, collisions are considered inelastic to match the qualitative behavior seen in experiment.

\section{Numerical Results}
\begin{figure}
	\includegraphics[width=8.6cm]{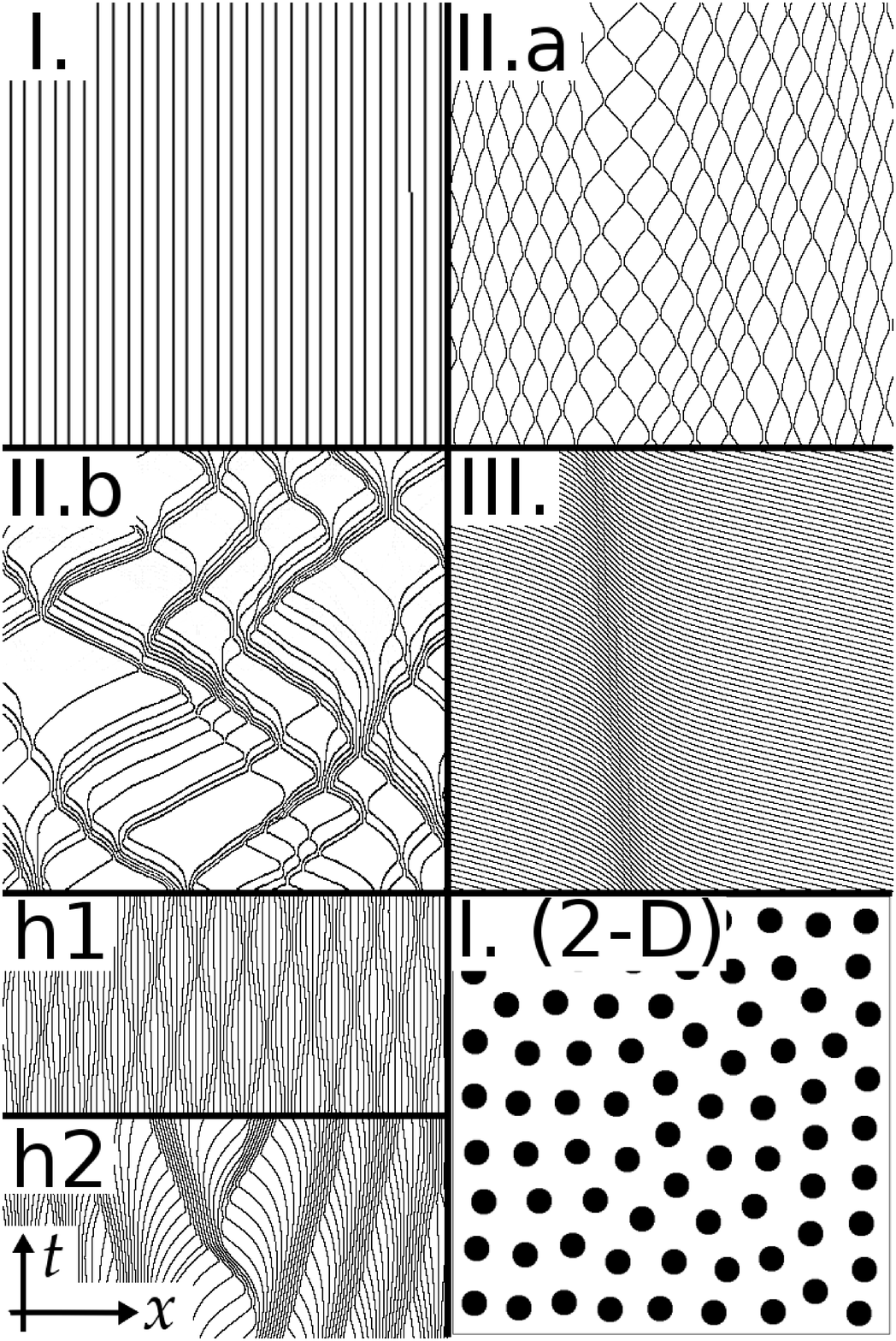}
	\caption{Space-time diagrams showing I) stationary phase, II.a) synchronized oscillation, II.b) erratic oscillation, III) unidirectional flow (boat density = 0.2). Note the left moving dense region in III) representing a jam. h1 and h2 are higher density versions of II.a and II.b respectively(density = 0.6). The lower right figure shows a 2-D stationary state.}
	\label{fig:patterns}
\end{figure}

In 1-D numerical simulations, three distinct categories of behavior were observed. Listed in order of decreasing viscosity, these are: \bf (I)\rm stationary equilibrium, \bf (II)\rm oscillation, \bf (III)\rm unidirectional flow. Examples of these patterns are shown in the space-time diagrams of Fig.(\ref{fig:patterns}). The parameter determining the mode of behavior is the viscosity of the water, $\mu$ in the equations above. It also depends on the total density of boats as described below. 2-D simulations were also performed, and some observations are described, but the behavior of interest was more clearly demonstrated in 1-D.

An order parameter for the transition between stationary and oscillating phases is the ensemble average of the root mean square velocity, $\sqrt{\left<v^{2}\right>}$, where the average inside the root is taken over time. For the transition between oscillating and flowing phases, the order parameter is the average velocity, $\lvert \left<v\right>\rvert$, which is also averaged over both time and the ensemble. The time averages are taken over a long interval($\Delta t>800$), corresponding to hundreds of oscillation periods and beginning after an initial relaxation period. The characteristic oscillation period depends on boat density and viscosity. The shortest periods were approximately 1 unit of time, but some irregular oscillations close to the flow transition took more than 10 units and required longer simulation intervals. The initial positions were semi-randomized to break symmetry. 

We will now qualitatively describe the three different types of behavior, enumerated as above, and the transitions between them occurring at the critical viscosities $\mu_{c1}$ and $\mu_{c2}$.

\bf (I).\rm In the very high viscosity phase, the system approaches a stable, stationary equilibrium in which the boats are uniformly spaced. 2-D simulations also show a stationary, crystal-like pattern which very closely resembles the experimental observations by Soh, Bishop and Grzybowski~\cite{Soh2008}. Furthermore, a transition between moving and stationary states occurs depending on the total density of boats, agreeing with experiment. This consistency suggests that the behavior seen in experiment may be influenced by surface tension rather than purely hydrodynamic effects.
\begin{figure}
	\includegraphics[width=8.6cm]{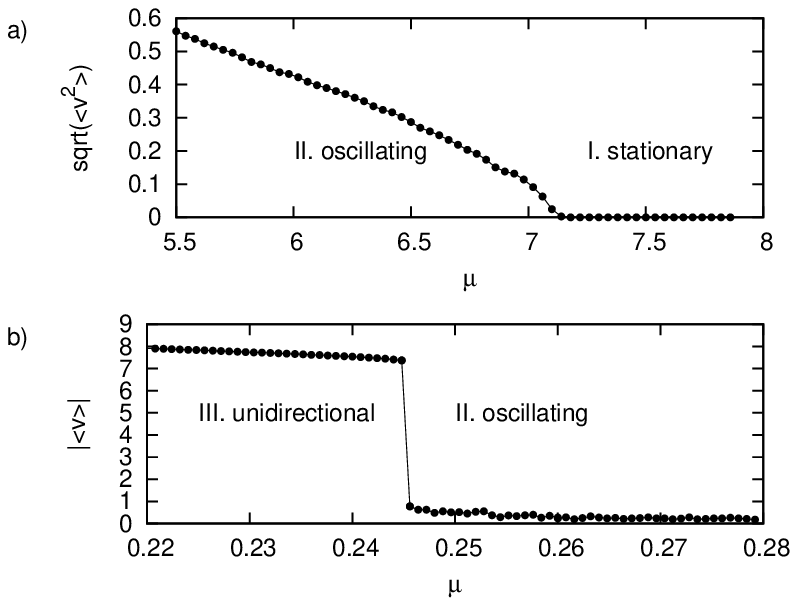}
	\caption{a) $\sqrt{\left<v^{2}\right>}$ in the vicinity of $\mu_{c1}$. b) $\lvert \left<v\right>\rvert$ in the vicinity of $\mu_{c2}$. Other parameters are: $k=0.072$, $\Gamma=528$, $\alpha=7.2$, $r_{0}=0.25$, $R=75$, $N=20$}
	\label{fig:velocity}
\end{figure}

\bf (II).\rm As the viscosity is decreased beyond $\mu_{c1}$, the stationary state becomes unstable and the boats begin to move. This onset of motion can be quantified by an abrupt increase from zero in $\sqrt{\left<v^{2}\right>}$ as shown in Fig.\ref{fig:velocity}.a. For density greater than about $0.15$, the boats oscillate and $\lvert \left<v\right>\rvert$ remains almost zero.  For lower densities, $\lvert \left<v\right>\rvert$ depends highly on the initial conditions, but the change in $\sqrt{\left<v^{2}\right>}$ is similar. The 2-D motion depends on the configuration of boats and clear oscillation was not seen. There are two different oscillation patterns depending on viscosity, labeled here as II.$a)$ and II.$b)$.

\begin{figure}
	\includegraphics[width=8.6cm]{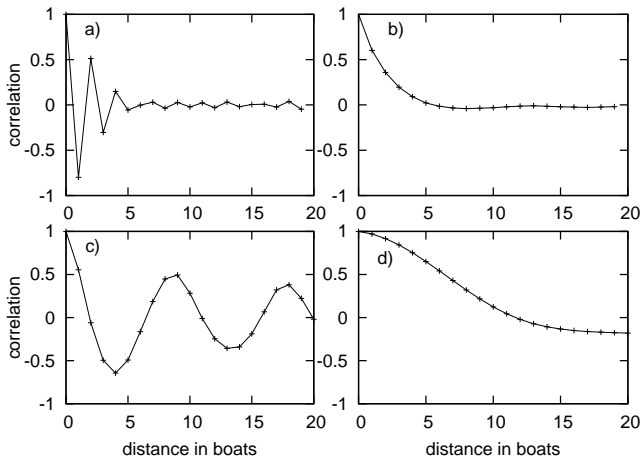}
	\caption{Cross correlation of velocity vs. distance in boat numbers for a) density=0.26, $\mu=0.6$ b) density=0.26, $\mu=4.0$ c) density=0.6, $\mu=0.12$ d) density=0.6, $\mu=0.68$ corresponding to behavior shown in Fig.\ref{fig:patterns} II.$a$, II.$b$, h1, h2 respectively}
	\label{fig:correlations}
\end{figure}

II.$a)$ For higher viscosity, the boats oscillate in a very synchronized formation as shown in Fig.\ref{fig:patterns}. To quantify this, we can measure the degree of synchronization between boats using the cross correlation of their velocities. Fig.\ref{fig:correlations}.a,c shows the correlation vs. the distance in numbers of boats. If the density is below about $0.4$, neighboring boats are anti-synchronized. Higher densities show the strongest anti-synchronization at a distance of several boats. This distance appears to increase with increasing total density.

II.$b)$ For lower viscosity, the oscillations become more irregular and the synchronized behavior vanishes. As seen in Fig.\ref{fig:correlations}.b,d, the  correlation is positive in the vicinity of the boats and approaches zero for boats further away. There is no pattern of synchronization and anti-synchronization as seen for type II.$a)$ behavior.

\bf (III).\rm As viscosity is decreased further, the boats move in larger groups and with longer periods between changes in direction. There is a critical viscosity, $\mu_{c2}$, below which the boats no longer change direction and the flow becomes unidirectional. The selection of direction is spontaneous and depends on initial conditions. Quantitatively, there is a discontinuity in the net velocity of the entire system, $\lvert \left<v\right>\rvert$ as shown in Fig.\ref{fig:velocity}.b. To test for hysteresis in this discontinuity, the viscosity was gradually increased and decreased using a few different values of density. For the parameter space tested, hysteresis was not observed. This phase can exhibit several different kinds of flow including free flow, jammed flow and pulsing flow. Different types of flow have been studied elsewhere~\cite{Suematsu20102}, and they are not distinguished here.

\begin{figure}
	\includegraphics[width=8.6cm]{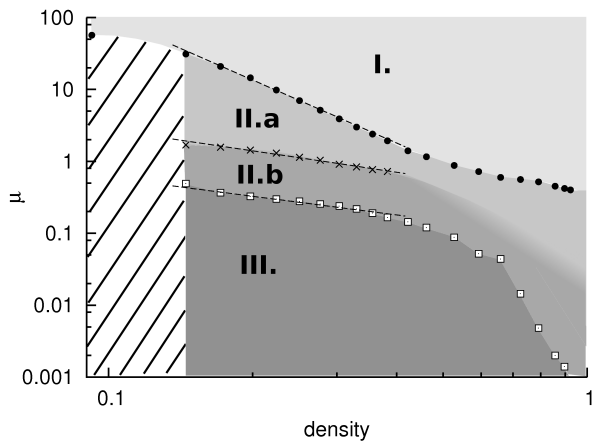}
	\caption{Phase diagram  for the types of behavior. The dashed lines are a power law fit for densities between 0.17 and 0.38, but such a relation is not implied.(same parameters as fig.\ref{fig:velocity})}
	\label{fig:critvals}
\end{figure}

Fig.\ref{fig:critvals} is a phase diagram illustrating the regions of viscosity and density for each type of collective behavior. A power law fit was made for the density range 0.17 to 0.38, but it is not clear that the data follow such a relation, especially in the case of $\mu_{c2}$. Although the critical values $\mu_{c1}$ and $\mu_{c2}$ are clearly defined, the change between patterns II.$a$ and II.$b$ is more gradual. The values shown correspond to the point at which the average correlation between neighboring boats increases to zero. At higher density the change was too gradual to assign a meaningful transition point, so it is not clearly defined in the diagram. The non-stationary phase for density below about $0.15$ depends highly on initial conditions, so the distinction between oscillating and flowing phases is not shown. However, the transition between stationary and moving phases remains even down to the low density limit of one boat.

The transition at $\mu_{c1}$ and the change from pattern II.$a$ to II.$b$ appear to be independent of the total system size as long as the density is kept constant. The value of $\mu_{c2}$ increases very slightly with increasing system size. This was tested by varying the total route length between $R=38$ and $R=607$.

\section{Analysis}
\begin{figure}
	\includegraphics[width=8.6cm]{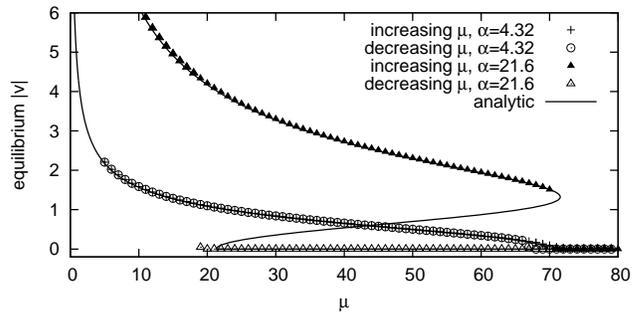}
	\caption{Equilibrium velocity for $N=1$.}
	\label{fig:oneboat}
\end{figure}

The $\mu_{c1}$ transition can be analyzed for one boat by assuming that it has reached an equilibrium speed, and the camphor field is time independent in the frame of the boat. $c$ will then satisfy a traveling pulse equation;
\begin{equation}
-v \nabla c = \Delta c - kc + \alpha F(x-x_{0})
\label{eq:oneboat_c} 
\end{equation}
Nagayama, Nakata, Doi and Hayashima have provided a detailed analysis of this case for a single camphor pellet in 1-D~\cite{Nagayama2004}. They demonstrated a bifurcation between a stable stationary solution and a constant-velocity solution for a camphor pellet on water. Also, the bifurcation is supercritical for small $r_{0}$, and subcritical for large $r_{0}$. The treatment for a CB is the same, but depends on $c$ at the edge of the boat rather than the pellet, so bifurcation type can also depend on the source rate, $\alpha$. The details of this analysis are given in the appendix. The results of analytical calculations and simulation are shown in Fig.\ref{fig:oneboat}. This result can be extended to a uniformly spaced ensemble of boats by shrinking the periodic domain to include exactly one boat. Unfortunately, it is only valid if all boats are moving identically, which does not describe the oscillatory behavior observed. In 2-D the trajectory of the boat may be irregular as was shown in experiment~\cite{Soh2008}, so this technique is not directly applicable.

To explain the oscillatory behavior, there must be an effective repulsive force acting on boats that are close together. This can be understood by constructing a simple scenario. First assume that two boats in the system begin moving toward each other, and that their initial separation is sufficient that they can approach equilibrium velocity before colliding. Because their speeds are nearly equal, and the collisions are modeled as inelastic, the boats will nearly come to a stop after colliding. Also, assume that the relaxation time of the local camphor field is sufficiently short. The camphor field will have the form of two neighboring peaks with exponentially decreasing sides. Due to the contribution of each camphor source, the concentration directly between the peaks, corresponding to the closer edges of the boats, will be higher than that at the further edges of the boats. The higher surface tension on the further edges will provide an effective repulsion. The repulsion does not need to have a long range, because once a boat is moving in one direction, it will tend to continue in that direction. This process describes the regular oscillation seen in II.$a$. However, if the velocities of the two boats are significantly different, the final velocity after colliding may be large enough that both boats will begin traveling in the same direction. The exact speed difference required would depend sensitively on the camphor profile and system parameters. This sensitivity may explain the seemingly chaotic oscillation patterns seen in II.$b$.

The transition between flow and oscillation cannot be treated by such simplifications because the collective behavior is extremely complicated and collision effects are significant. Instead, we will outline a reasonable scenario based on observations. Consider a single boat in the ensemble. The camphor field influencing it can be approximated as a combination of the boat's local field generated by its camphor pellet and a large-scale field generated by the ensemble. The local part typically has the shape of a sharp, localized peak, while the large-scale part is much smoother and varies over the entire route. Using these two parts, the driving force can then be separated into local and large-scale parts. The local part typically propels the boat in the direction of its velocity, while the large-scale part may act in either direction depending on the structure of $c$. We will define $f_{l}$ as the component of the force generated by a boat's local field. $f_{g}$ is the component generated by the large-scale field.

Assume that $f_{l}$ is constant, and that $f_{g}(x,t)$ depends on the large-scale profile of $c(x,t)$ which may change in time. Then the condition $f_{l}+f_{g}(x,t)<\mu v$ will cause the boat to slow down. Recall that the driving force depends on the difference in $c$ between the two sides of the boat and acts in the direction of decreasing concentration. This means that if a boat is moving toward an increasing gradient in $c$, $f_{g}$ will be negative and the boat may slow down. Such a condition is often created by a boat density peak such as a jam, which generates a peak in $c$. In the flow phase, the boat's momentum is enough to overcome the opposing gradient as it approaches the peak. If the viscosity is increased, the boat will have a lower speed, and may stop and reverse direction before reaching the peak. This reversal happens repeatedly for viscosity above $\mu_{c2}$, causing oscillation. Note that the boats may pass the peak several times or collide with other boats before finally reversing direction.

In summary, we have theoretically investigated a new kind of self-propelled particle and found several distinct patterns of collective motion and two kinetic phase transitions. The transitions can be quantified by abrupt changes in the flow, $\lvert \left<v\right>\rvert$, and the root mean square velocity, $\sqrt{\left<v^{2}\right>}$ of all the boats in the system. We have described the different patterns of self organized behavior seen in numerical simulations, and outlined a brief analysis of the transitions. In future work, we hope to test these results through experiment.

\begin{acknowledgments}
This work is supported by Grants-in-Aid for Scientific Research(No.22540391) to H.N. and for Young Scientist(B)(No.23740299) to N.J.S. from the Ministry of Education, Science and Culture of Japan and the Global COE Program Formation and Development of Mathematical Sciences Based on Modeling and Analysis.
\end{acknowledgments}

~\bibliography{camphorRefs}

% appendix
\appendix
\section{}
The camphor field created by one boat moving at constant velocity is described by the traveling pulse equation.
\begin{equation}
-v \nabla c = \Delta c - kc + \alpha F(x-x_{0})
\label{eq:apponeboat_c} 
\end{equation}
The following solution is adapted from Nagayama, Nakata, Doi and Hayashima ~\cite{Nagayama2004} It assumes that the route length is sufficiently long that the profile of $c$ is well localized. This allows us to adopt the boundary condition $\lim_{\lvert x \rvert \rightarrow \infty }c(x) = 0$ and continuity condition, $c(x) \in C^{1}(\Re)$. Also, we will use the slightly different dimensionless parameters $c = \frac{c}{\alpha}$ and $\beta = \beta\alpha$ to simplify the calculation. The solution is
\begin{equation}
c(x)=\left\{
\begin{array}{lr}
	A_{1}\exp \left( \frac{1}{2}\eta_{+}x \right) & x<-r_{0} \\
	A_{2}\exp \left( \frac{1}{2}\eta_{+}x \right) + B_{2}\exp \left( \frac{1}{2}\eta_{-}x \right) -\frac{1}{k} & -r_{0}<x<r_{0} \\
	B_{3}\exp \left( \frac{1}{2}\eta_{-}x \right) & x>r_{0}
\end{array}
\right.
\label{eq:csolution} 
\end{equation}
where
\begin{equation}
\begin{array}{l}
	A_{1} = \frac{\eta_{-}}{2k\eta}\left( \exp \left(-\frac{\eta_{+}}{2}r_{0}\right) - \exp \left(\frac{\eta_{+}}{2}r_{0}\right) \right) \\
	A_{2} = \frac{\eta_{-}}{2k\eta} \exp \left(-\frac{\eta_{+}}{2}r_{0}\right) \\
	B_{2} = -\frac{\eta_{+}}{2k\eta} \exp \left(\frac{\eta_{-}}{2}r_{0}\right) \\
	B_{3} = \frac{\eta_{+}}{2k\eta}\left( \exp \left(-\frac{\eta_{-}}{2}r_{0}\right) - \exp \left(\frac{\eta_{-}}{2}r_{0}\right) \right) \\
	\eta = \sqrt{v^{2}+4k} \\
	\eta_{\pm} = -v \pm \eta
\end{array}
\label{eq:csolutionconstants} 
\end{equation}
The values of this solution at the two edges of the boat, $x=\pm\frac{L}{2}=\pm2r_{0}$ are then
\begin{equation}
\begin{array}{l}
	c_{+} = c(2r_{0}) = B_{3}\exp \left( \eta_{-}r_{0} \right) \\
	c_{-} = c(-2r_{0}) = A_{1}\exp \left( -\eta_{+}r_{0} \right)
\end{array}
\label{eq:catthesides} 
\end{equation}
Substituting these and the constant velocity, $v$, into the equation of motion gives
\begin{equation}
0 = -\mu v +\Gamma \left[ \frac{1}{\beta B_{3}^{2}\exp \left( 2\eta_{-}r_{0} \right) + 1} - \frac{1}{\beta A_{1}^{2}\exp \left( -2\eta_{+}r_{0} \right) + 1} \right]
\label{eq:motionsolution}
\end{equation}
Note that if $v=0$, $A_{1}=B_{3}$ and $\eta_{+}=-\eta_{-}$, so the equation of motion is satisfied for any value of $\mu$. We can also rearrange the equation to give viscosity as a function of equilibrium velocity.
\begin{equation}
\mu(v) = \frac{\Gamma}{v} \left[ \frac{1}{\beta B_{3}^{2}\exp \left( 2\eta_{-}r_{0} \right) + 1} - \frac{1}{\beta A_{1}^{2}\exp \left( -2\eta_{+}r_{0} \right) + 1} \right]
\label{eq:muofv}
\end{equation}
This was used to produce the analytical solutions shown in Fig.\ref{fig:oneboat}. Inverting this relationship results in a bifurcation as shown in the figure. By changing $\beta$ in this equation, the bifurcation changes between subcritical and supercritical. Recalling the different dimensionless parameters used here, this is equivalent to changing $\alpha$ for the parameters used in the main part of the text.

\end{document}